\documentclass[reprint, amsmath, amssymb, aps, superscriptaddress]{revtex4-1}
\usepackage{format}
\usepackage{soul} 
\allowdisplaybreaks

\begin{document}

\preprint{APS/123-QED}

\title{ Simulations of evaporation to deep Fermi degeneracy in microwave-shielded molecules }
  
\author{Reuben R. W. Wang}
\affiliation{JILA, NIST, and Department of Physics, University of Colorado, Boulder, Colorado 80309, USA}
\author{Shrestha Biswas}
\author{Sebastian Eppelt}
\affiliation{Max-Planck-Institut f\"{u}r Quantenoptik, 85748 Garching, Germany}
\affiliation{Munich Center for Quantum Science and Technology, 80799 M\"{u}nchen, Germany}
\author{Fulin Deng}
\affiliation{School of Physics and Technology, Wuhan University, Wuhan, Hubei 430072, China}
\affiliation{CAS Key Laboratory of Theoretical Physics, Institute of Theoretical Physics, Chinese Academy of Sciences, Beijing 100190, China}
\author{Xin-Yu Luo}
\affiliation{Max-Planck-Institut f\"{u}r Quantenoptik, 85748 Garching, Germany}
\affiliation{Munich Center for Quantum Science and Technology, 80799 M\"{u}nchen, Germany}
\author{John L. Bohn}
\affiliation{JILA, NIST, and Department of Physics, University of Colorado, Boulder, Colorado 80309, USA}

\date{\today} 

\begin{abstract}

In the quest toward realizing novel quantum matter in ultracold molecular gases, we perform a numerical study of evaporative cooling in ultracold gases of microwave-shielded polar fermionic molecules. Our Monte Carlo simulations incorporate accurate two-body elastic and inelastic scattering cross sections, realistic modeling of the optical dipole trap, and the influence of Pauli blocking at low temperatures.  The simulations are benchmarked against data from evaporation studies performed with ultracold NaK molecules, showing excellent agreement.  We further explore the prospects for optimizing the evaporation efficiency by varying the ramp rate and duration of the evaporation trajectory. 
Our simulation shows that it is possible to reach $< 10\%$ of the Fermi temperature under optimal conditions even in the presence of two-body molecular losses. 

\end{abstract}

\maketitle

\section{\label{sec:introduction} Introduction}

Ultracold polar molecules have emerged as a clean and highly controllable platform for studying quantum chemistry \cite{Bohn17_Sci, Liu22_ARPC, Karman24_NatPhys} and dipolar quantum many-body physics \cite{Yan13_Nat, Hazzard14_PRL, Li23_Nat, Carroll24_arxiv}. Recent developments in collisional shielding \cite{Quemener16_PRA, Gonzalez17_PRA, Matsuda20_Sci, Li21_Nat, Lassabliere22_PRA, Avdeenkov12_PRA, Karman18_PRL, Karman20_PRA, Anderegg21_Sci, Lin23_PRX, Bigagli23_NatPhys} 
have enabled efficient evaporative cooling of polar molecules, leading to the creation of both degenerate Fermi gases and a Bose-Einstein condensate of polar molecules \cite{Valtolina20_Nat, Schindewolf22_Nat, Bigagli24_Nat}. This paves the way to explore novel quantum phases, including $p$-wave superfluids and the extended Fermi-Hubbard model in fermionic molecules \cite{Cooper2009, Bruun_stripephase_2012}, as well as quantum droplets, exotic supersolids, and Wigner crystals in bosonic molecules \cite{Schmidt2021}. 

The lowest reported temperature achieved for a Fermi gas of polar molecules is so far 0.36 times the Fermi temperature \cite{Schindewolf22_Nat}, substantially above the critical temperature required for realizing $p$-wave superfluidity in polar molecules and a Bose-Einstein condensate of tetratomic molecules \cite{Deng23_PRL}. Evaporative cooling of fermions to deep quantum degeneracy is hindered by Pauli blocking, which suppresses thermalizing elastic collisions \cite{Wang21_PRA} and introduces hole heating that becomes more severe under deeper quantum degeneracy \cite{mckay2011cooling}. Additionally, two-body collisional losses and rapid dipolar elastic collisions further limit the evaporative cooling efficiency in molecular Fermi gases compared to their atomic counterparts \cite{Schindewolf22_Nat}.

\begin{figure}[ht]
    \centering
    \includegraphics[width=\columnwidth]{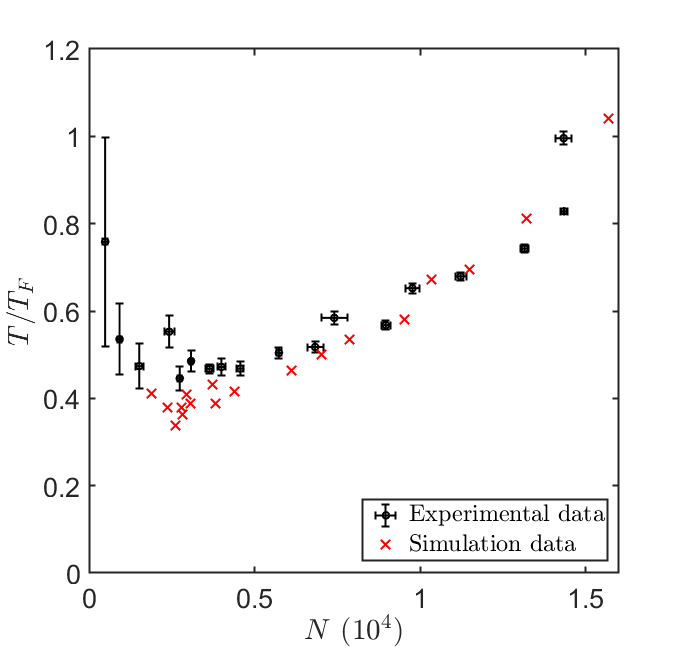}
    \caption{ Plot of the ratio of the gas to Fermi temperatures $T/T_F$, after evaporating down to $N$ molecules in experiments (black circles with error bars) with different final trap depths, compared to those obtained from our numerical simulations (red crosses). See the main text for further details. }
    \label{fig:ToTF_vs_N_comparison}
\end{figure}

In this work, we develop a comprehensive theoretical model to simulate, understand, and enhance the evaporative cooling of fermionic molecular gases.  
Our simulations are capable of recreating the observed evaporation trajectories with microwave-shielded $^{23}$Na$^{40}$K molecules down to quantum degeneracy, recently achieved in Ref.~\cite{Schindewolf22_Nat}.
In that study, the final trap depth was varied over several experimental instances with forced evaporation occurring over $150$ ms, resulting in various molecule numbers and gas temperatures attained at the end of each evaporation trajectory. 
Fig.~\ref{fig:ToTF_vs_N_comparison} showcases the favorable agreement of our numerically simulated data points (red crosses) with the experimental measurements (black circles with error bars), except for disagreement at low molecule numbers attributed to experimental trap jitter at low laser powers. 
The purpose of this paper is, therefore, to provide details of our simulation methods, survey parameters relevant to current evaporation experiments, and open avenues for broader evaporation studies for collisionally-shielded polar molecules. 

The rest of this paper is organized as follows.
Sec.~\ref{sec:formulation} presents the theoretical model we utilize for our simulations, with details of its numerical solutions given in Sec.~\ref{sec:MonteCarlo_solver}. The procedure for forced evaporation is discussed in Sec.~\ref{sec:evap_simulations}, along with the observables we extract from our solver that mimic those of experiments. 
Recommendations for efficient cooling from simulation results are presented in Sec.~\ref{sec:numerical_results}, following which pathways to going below $10\%$ of the Fermi temperature are explored in Sec.~\ref{sec:superfluid_prospects}. Final remarks and conclusions are drawn in Sec.~\ref{sec:conclusion}.

\section{ \label{sec:formulation} Kinetic Fermi Gases }

At nonzero temperatures, the dynamics of  ultracold Fermi gases are well described by the quantum Boltzmann equation, which provides a statistical description of rarefied gases in phase space \cite{Bonasera94_PR}. These many-body systems are collectively described by a single-particle phase space distribution function $f(\boldsymbol{r}, \boldsymbol{p}, t)$, defined such that
\begin{align} \label{eq:normalization} 
    \int d^3 r d^3 p f(\boldsymbol{r}, \boldsymbol{p}, t) = N,
\end{align}
with phase space position and momentum coordinates $\{\boldsymbol{r}, \boldsymbol{p}\}$. The Boltzmann equation is then written as \cite{Reif09_Waveland}
\begin{align} \label{eq:Boltzmann_equation} 
    \bigg(
    \frac{ \partial }{ \partial t }
    & +
    \frac{ \boldsymbol{p} }{ m }
    \cdot 
    \grad_r 
    -
    \grad_r V(\boldsymbol{r}) \cdot \grad_p
    \bigg)
    f(\boldsymbol{r}, \boldsymbol{p}, t)
    =
    \mathcal{I}[f],
\end{align}
where $m$ is the molecular mass, $V(\boldsymbol{r})$ is the trap potential and $N$ is the number of molecules.

Molecular collisions are accounted for via the collision integral ${\cal I}[f]$.  
With the goal of cooling to deep quantum degeneracy, Pauli blocking must be included in the kinetic model for temperatures lower than the Fermi temperature $T_F$ 
\cite{Nordhiem1928_PRSL, Uehling33_PR}:
\begin{align} \label{eq:collision_integral_fermions}
    \mathcal{I}[f]
    &=
    \int \frac{d^3 p_1}{m} \abs{\boldsymbol{p} - \boldsymbol{p}_1} 
    \int d\Omega' \frac{d\sigma}{d\Omega'} \nonumber\\
    &\quad\quad \times
    \big[
    f' f'_1 
    \left( 1 - h^3 f \right) 
    \left( 1 - h^3 f_1 \right) \nonumber\\
    &\quad\quad\quad -
    f f_1 
    \left( 1 - h^3 f' \right) 
    \left( 1 - h^3 f'_1 \right)
    \big],    
\end{align}
where ${d\sigma / d\Omega}$ is the differential cross section, and we use the shorthand notation $f_1 = f(\boldsymbol{r}, \boldsymbol{p}_1, t)$ and $f' = f(\boldsymbol{r}, \boldsymbol{p}', t)$. 
The Pauli blocking factors $-h^3 f$ become significant at phase space densities (PSD) of order $\rho_{\rm PSD} \gtrsim 0.1$, where PSD is defined as
\begin{align} \label{eq:PSD}
    \rho_{\rm PSD}
    &=
    \langle n \rangle
    \lambda_{\rm th}^3,
\end{align}
which compares the ensemble averaged number density $\langle n \rangle$ against the cubed thermal de Broglie wavelength $\lambda_{\rm th} = h / \sqrt{ 2 \pi m k_B T }$.  Above, $h$ is Planck's constant, $k_B$ is Boltzmann's constant, and $T$ is the gas temperature.

Our study considers a single species gas of $N$ dipoles, aligned along the dipole polarization axis $\boldsymbol{{\rm E}}$. The gas is confined in a crossed optical dipole trap (xODT) with 2 perpendicular intersecting Gaussian beams, modeled by the potential
\begin{align} \label{eq:trap_potential}
    V(\boldsymbol{r}) 
    &=
    V_{\rm ODT}(\boldsymbol{r}) 
    +
    V_{\rm g}(\boldsymbol{r}),
\end{align}
with terms above defined as \cite{Grimm00_AAMOP}
\begin{subequations}
\begin{align}
    V_{\rm ODT}(\boldsymbol{r})
    &=
    - \frac{ 2 \alpha_1 P_1 \exp\left( 
    -\frac{ 2 y^2 }{ w_{1,y}^2(x) } 
    -
    \frac{ 2 z^2 }{ w_{1,z}^2(x) } 
    \right) }{ \pi w_{1,y}(x) w_{1,z}(x) } \nonumber\\
    &\quad -  
    \frac{ 2 \alpha_2 P_2 \exp\left( 
    -\frac{ 2 x^2 }{ w_{2,x}^2(y) } 
    -
    \frac{ 2 z^2 }{ w_{2,z}^2(y) } 
    \right) }{ \pi w_{2,x}(y) w_{2,z}(y) }, \\ 
    V_{\rm g}(\boldsymbol{r})
    &=
    m g z.
\end{align}
\end{subequations}
Above, $P_i$ is each beam's laser power, $\alpha_i$ the molecular polarizability given the wavelength of beam $i$, 
\begin{align}
    w_{\mu, i}(r) 
    =
    W_{\mu, i} \sqrt{ 1 + \frac{ r^2 }{ R_{\mu, i}^2} },
\end{align}
with $R_{\mu, i} = \pi W_{\mu, i}^2 / \lambda$ denoting the Rayleigh length and $W_{\mu, i}$ is the beam width of wavelength $\lambda$.
Altering the power of the xODT lasers can alter the trap depths, allowing molecules to spill out due to gravity and produce evaporative cooling. 
At high laser powers, $V_{\rm ODT}$ is well approximated as harmonic around the trap minima:
\begin{align}
    V_{\rm harm}(\boldsymbol{r})
    &=
    \frac{1}{2} m \sum_{\nu} \omega_\nu^2 r_\nu^2,
\end{align}
where the gas mostly resides. Above, $\omega_\nu$ are the harmonic trap frequencies along coordinate axis $\nu$ defined in terms of the laser parameters as detailed in App.~\ref{app:harmonic_trap}.

\section{ \label{sec:MonteCarlo_solver} Monte Carlo Solutions }

The existing literature on numerical solutions to the Boltzmann equation provides a strong foundation for this work \cite{Bird70_PF, Goulko12_NJP, Pantel15_PRA, Sykes15_PRA, Wang20_PRA}. For completeness, we briefly present our implementation of these methods in this section.  

Numerical solutions to Eq.~(\ref{eq:Boltzmann_equation}) are performed by stepping forward in time with discrete time steps. In our case, there are two main features of the dynamics: 1) free-stream evolution influenced by the trapping potential $V(\boldsymbol{r})$ [left-hand side of Eq.~(\ref{eq:Boltzmann_equation})] and 2) two-body collisional interactions [right-hand side of Eq.~(\ref{eq:Boltzmann_equation})], motivating the definition of two distinct time scales:
\begin{subequations}
\begin{align}
    \quad \tau_V 
    &=
    \frac{ 2\pi }{ \min_{\nu} \{ \omega_\nu \} }, \\
    \tau_\mathrm{coll} 
    &= 
    \langle n \sigma v_r \rangle^{-1}, 
\end{align}
\end{subequations}
where $\sigma$ is the total cross section, $v_r$ is the relative velocity and $\langle \ldots \rangle$ denotes a molecular ensemble average. 
We take advantage of this distinction by time evolving each physical process with its own time step, free-stream kinetics with $\Delta t \ll \tau_V$, and collisions with $\delta t \ll \tau_{\rm coll}$. Furthermore, $\delta t$ is taken to be adaptive \cite{Frenkel01_El}, updated based on the mean collision rate at any given time in the simulation.

\subsection{ \label{sec:classical_evolution} Free-stream kinetics }

Direct solutions of the 6-dimensional phase space distribution are, in general, extremely expensive to solve numerically. 
Instead, we adopt the approximation employed by Ref.~\cite{Bird70_PF}, where $f(\boldsymbol{r}, \boldsymbol{p})$ is discretized into phase space points we refer to as ``simulation particles":
\begin{align} \label{eq:phasespace_discretized}
    f(\boldsymbol{r}, \boldsymbol{p}) 
    \approx
    \xi \sum_{k = 1}^{N_{\rm sim}} \delta^3(\boldsymbol{r} - \boldsymbol{r}_k) \delta^3(\boldsymbol{p} - \boldsymbol{p}_k), 
\end{align}
where $\xi = N/N_{\rm sim}$ is the ratio of the actual number of particles $N$, to simulation particles $N_{\rm sim}$. 
For most considerations here, we set $N_{\rm sim} = N$ so that each evaporated particle depicts the expected cooling effect from such molecular loss.  
Each test particle then evolves under Newton's equations, which is numerically performed with the Verlet symplectic algorithm \cite{Verlet67_PR}:
\begin{subequations} \label{eq:Verlet_integration}
\begin{align}
    \boldsymbol{q}_k 
    &=
    \boldsymbol{r}_k(t) 
    +
    \frac{\Delta t}{2m}
    \boldsymbol{p}_k(t), \label{eq:Verlet_step1} \\
    \boldsymbol{p}_k(t + \Delta t) 
    &= 
    \boldsymbol{p}_k(t) 
    +
    \boldsymbol{F}_k\Delta t, \label{eq:Verlet_step2} \\
    \boldsymbol{r}_k(t + \Delta t) 
    &= 
    \boldsymbol{q}_k 
    +
    \frac{\Delta t}{2m}
    \boldsymbol{p}_k(t + \Delta t), \label{eq:Verlet_step3}
\end{align}
\end{subequations} 
where subscripts $k$ denote the $k$-th test particle, $t$ is the current time and $\boldsymbol{F}_k = -\grad V_{\rm ODT}(\boldsymbol{r})$.

\subsection{ \label{sec:collision_integral} Quantum collision integral }

The collision integral is computed via a direct simulation Monte Carlo (DSMC) method \cite{Bird70_PF} at time intervals $\delta t$. The DSMC method utilizes a discrete spatial grid for efficiency, where we adopt a locally adaptive discretization scheme to construct it based on local density variations.

\begin{figure}[ht]
    \centering
    \includegraphics[width=\columnwidth]{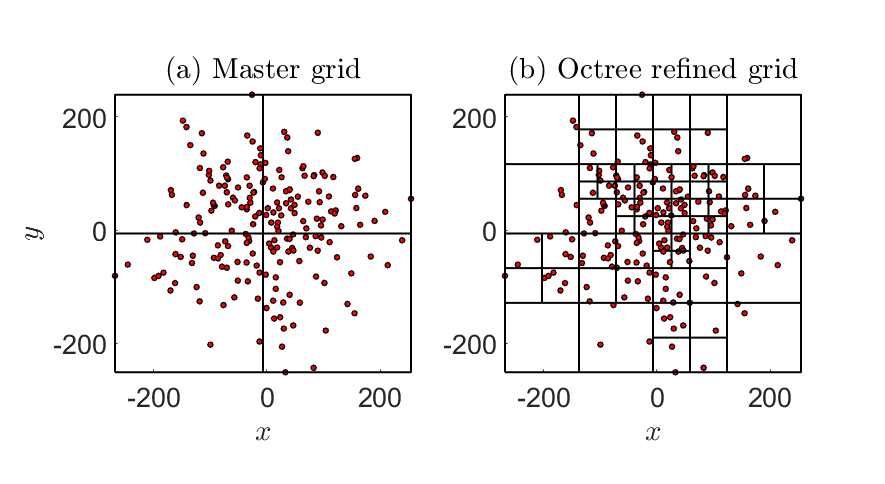} 
    \caption{ Visualization of the octree algorithm refinement (b) applied to the master grid (a), projected in 2-dimensions. In this figure, there are initially 8 master grid cells with $N = 200$, and $N_{\text{cell}}^{\max} = 3$, with Gaussian distributed points in arbitrary units.  }
    \label{fig:OcTree_example}
\end{figure}

The spatial grid is built at every $\delta t$ time step in two phases. Phase one is the construction of a master grid, consisting of uniform volume cells that span the simulation volume surrounding the test particle ensemble. 
From this, the resolution of the spatial grid is then refined in phase two with an octree algorithm \cite{Franklin85_Springer}. The algorithm recursively subdivides each master grid cell into eight octants, terminating only when each cell has at most $N_{\rm cell}^{\max}$ test particles. 
The parameter $N_{\rm cell}^{\max}$, is initialized at the start of the simulation and can be optimized for stochastic convergence. The octree refinement of an initial master grid is illustrated in Fig.~\ref{fig:OcTree_example}. 
In practice, we typically use $N_{\rm cell}^{\max} = 5$ to $10$.

A dilute gas allows us to consider collisions as occurring only within grid cells, resulting in their exiting from a phase space volume element $d^3 r d^3p$, at a rate given by $d^3 r d^3 p {\cal I}^{\rm out}[f]$, 
where
\begin{align}
    {\cal I}^{\rm out}[f] 
    &= 
    - \int \frac{d^3 p_1}{m} \abs{\boldsymbol{p} - \boldsymbol{p}_1} 
    \int d\Omega' \frac{d\sigma}{d\Omega'} \nonumber \\
    &\quad\quad\quad\quad\quad \times 
    f f_1 
    \left( 1 - h^3 f' \right) 
    \left( 1 - h^3 f'_1 \right).
\end{align}
In the simulation, we replace the product of distributions in a differential phase space volume $d^3 q d^3 p f f_1$, by all pairs of test particles in a grid cell along with their associated momenta $\boldsymbol{p}$ and $\boldsymbol{p}_1$. This facilitates Monte Carlo integration of the integral over $d^3 p_1$, with integrand 
\begin{align} \label{eq:collision_rate_integrand}
    \frac{ 1 }{ m } 
    \abs{\boldsymbol{p} - \boldsymbol{p}_1} f f_1 \int d\Omega' \frac{d\sigma}{d\Omega'} 
    \left( 1 - h^3 f' \right) 
    \left( 1 - h^3 f'_1 \right), 
\end{align}
to get a collision rate. Obtaining values of the integrand above requires further computation of the effective total cross section
\begin{align}
    \sigma_\mathrm{eff} 
    =
    \int d\Omega' \frac{d\sigma}{d\Omega'} 
    \left( 1 - h^3 f' \right) 
    \left( 1 - h^3 f'_1 \right),
\end{align}
which we also evaluate through Monte Carlo integration over $d\Omega'$. The two sequential integration steps above correspond to a sampling of: 1) collision occurrences, and 2) post-collision momenta. 

In step (1), a collision proceeds between test particles $i$ and $j$, with probability \cite{Jackson02_PRA, Sykes15_PRA}
\begin{align} \label{eq:collision_probability}
    P_{ij} 
    &\approx 
    \xi \frac{\delta t}{m \Delta V_\mathrm{cell}} 
    \abs{ \boldsymbol{p}_{ij} }
    \sigma 
    (\boldsymbol{p}_{i j}),
\end{align}
where $\boldsymbol{p}_{ij} = \boldsymbol{p}_i - \boldsymbol{p}_j$, and the total cross section is the sum $\sigma = \sigma_{\rm inel} + \sigma_{\rm el}$,
of the inelastic and elastic cross sections respectively.
Details on the inelastic cross sections with universal short-range loss are provided below.
Inelastic collisions are then sampled to occur with probability $\sigma_{\rm inel} / \sigma$, following which that pair of molecules is discarded from the simulation.
Otherwise, the algorithm proceeds to step (2).

In step (2), post-collision momenta are sampled \cite{Bird70_PF, Sykes15_PRA} from the anisotropic differential cross section, following the occurrence of an elastic collision. For this work, we adopt the differential cross section derived in Ref.~\cite{Bohn14_PRA} at threshold, although strictly speaking, the large dipole moment of NaK has evaporation starting generally slightly outside of the scattering threshold regime \cite{Chen23_Nat}. 
Moreover, the elastic cross sections are those appropriate to microwave-shielded molecules, rather than point dipoles, but the differential cross sections are identical in the threshold limit.
The simulation results here are therefore a little more optimistic than reality \cite{Wang24_PRR}, but compare favorably enough to experiments that we will leave inclusion of these non-threshold cross sections to a future work.  

Quantum statistics then requires an additional accept-reject step \cite{Goulko12_NJP}, where the sampled post-collision momenta are only accepted with probability 
\begin{align} \label{eq:Qstat_probability}
    P'_{ij} 
    =
    \left( 1 - h^3 f'_i \right) \left( 1 - h^3 f'_j \right),
\end{align}
otherwise, no collision is said to have occurred. 
Application of these steps to all test particles results in an approximation of the collision integral over $\delta t$.

\subsection{ Smoothing the particle distribution \label{subsec:distribution_interpolation} }

Sampling $f'$ in Eq.~(\ref{eq:Qstat_probability}) is problematic, since the particle distributions $f'_i$ and $f'_j$ are discretized in the simulation. We resolve this issue by ``smearing" the $\delta$-functions with a Gaussian convolution kernel \cite{Pantel15_PRA, Goulko12_NJP}. These kernels are taken to have spatial width $\varsigma_{\nu}$ along axis $\nu$ and momentum width $\varsigma_{p}$, such that discretization noise is smoothed out while the distribution function remains physically consistent. These criteria are encapsulated by the conditions \cite{Lepers10_PRA}
\begin{subequations} \label{eq:smoothing_conditions}
\begin{align}
    \varsigma_{p}
    \overline{\varsigma}_{q} 
    &\gg 
    h \xi^{1/3}
    \label{subeq:smoothing_lowerbound}
    \\
    \varsigma_{\nu}
    &\ll
    R_{\nu}
    \frac{ T }{ T_F }, 
    \label{subeq:smoothing_upperbound_r}
    \\
    \varsigma_{p} 
    &\ll 
    p_F \frac{ T }{ T_F },
    \label{subeq:smoothing_upperbound_p}
\end{align}
\end{subequations}
where 
$p_F = \sqrt{ 2 m E_F }$ is the Fermi momentum, $R_{\nu} = \sqrt{ 2 E_F/(m \omega_{\nu}^2) }$ are the Thomas-Fermi radii \cite{Giorgini08_RMP} 
and $E_F = k_B T_F = \hbar \overline{\omega} ( 6 N )^{1/3}$ is the Fermi energy \cite{Butts97_PRA}.
Bars above quantities denote geometric means. We use widths defined by the geometric means of these upper and lower bounds, multiplied by a free parameter $\beta$, which is adjusted for stochastic convergence:
\begin{subequations}
\begin{align}
    \varsigma_{\nu}
    &=
    \sqrt{ 
    \left( \frac{ \hbar }{ m \omega_{\nu} } \right)^{1/2}
    \left( R_{\nu} \frac{ T }{ T_F } \right)
    }, \\
    \varsigma_{p}
    &=
    \sqrt{ 
    \left( m \hbar \overline{\omega} \right)^{1/2}
    \left( p_F \frac{ T }{ T_F } \right)
    }.
\end{align}
\end{subequations}
The convolution kernels are then
\begin{subequations} \label{eq:smeared_distributions}
\begin{align}
    \delta^3 (\boldsymbol{r} - \boldsymbol{r}_k) 
    &\to 
    g_r(\boldsymbol{r} - \boldsymbol{r}_k) 
    \equiv 
    \prod_{\nu=1}^3 \frac{ e^{-(r - r_k)^2 / \varsigma_{\nu}^2} }{ \sqrt{ \pi \varsigma_{\nu}^2 } }, \\
    \delta^3 (\boldsymbol{p} - \boldsymbol{p}_k) 
    &\to 
    g_p(\boldsymbol{p} - \boldsymbol{p}_k) 
    \equiv 
    \prod_{\nu=1}^3 \frac{ e^{-(p - p_k)^2 / \varsigma_{p_{\nu}}^2} }{ \sqrt{ \pi \varsigma_{p_{\nu}}^2 } }, 
\end{align}
\end{subequations}
which when used in Eq.~(\ref{eq:phasespace_discretized}), constitute a continuous distribution for evaluating  Eq.~(\ref{eq:Qstat_probability}).

\begin{figure}[ht]
    \centering
    \includegraphics[width=\columnwidth]{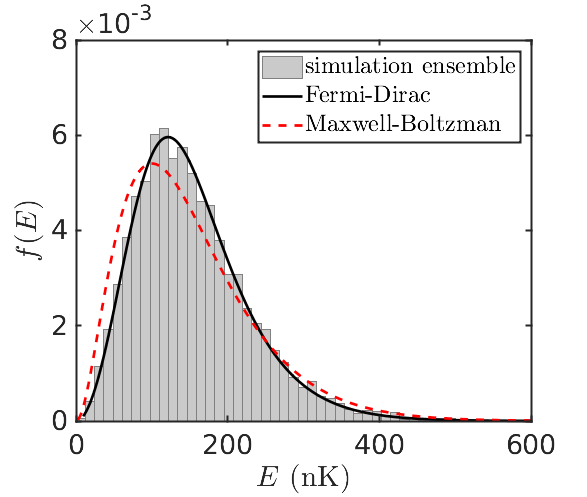}
    \caption{ Simulation ensemble energy distribution with $N = 10,000$ molecules (gray histogram), after collisional thermalization for $t = 0.5$ s from an intial Maxwell-Boltzmann distribution at $50$ nK (dotted red curve). The simulation achieves a Fermi-Dirac distribution (solid black curve) at $T \approx 42$ nK with a chemical potential of $\mu/k_B \approx 87$ nK. }
    \label{fig:FD_vs_MB_T=50nK_N=1e4}
\end{figure}

Including the Pauli blocking factors in Eq.~(\ref{eq:collision_integral_fermions}), enforces the equilibrium molecular distribution to obey Fermi-Dirac statistics, illustrated in Fig.~\ref{fig:FD_vs_MB_T=50nK_N=1e4}.
To obtain this plot, we ran an instance of the Monte Carlo simulation with $N = 10,000$ molecules in a perfectly harmonic trap of $(\omega_x, \omega_y, \omega_z) = 2\pi ( 45, 67, 157 )$ Hz, for a duration of $t = 0.5$ s. 
The molecules are initially sampled from a Maxwell-Boltzmann distribution (dashed red curve) at $T = 50$ nK ($T/T_F \approx 0.34$), then allowed to thermalize from elastic collisions to the gray histogram energy distribution. A Fermi-Dirac function \cite{Butts97_PRA} is then fitted to the histogram to obtain the solid black curve, with an extracted temperature of $T \approx 42$ nK and chemical potential $\mu/k_B \approx 87$ nK.
Fermi-Dirac statistic will affect how we perform thermometry during simulated evaporation, a topic we now turn to.

\section{ Simulations of dipolar evaporation \label{sec:evap_simulations}  }

During forced evaporation, the trap power is gradually reduced, lowering the trap depth which promotes the loss of energetic molecules. During this process, the trap powers follow the time-dependence
\begin{align} \label{eq:power_time_dependence}
    P_i(t)
    =
    P_i(0)
    -
    \Delta P_i
    \left(
    \frac{ 1 - e^{-t/\tau} }{ 1 - e^{-t_d/\tau} }
    \right),
\end{align}
where $\Delta P_i$ is the change in laser power of beam $i$, $t_d$ is the characteristic decay time and $\tau$ is the forced evaporation time. 
The log PSD versus log molecule number describes the forced evaporation trajectory, from which we can extract an evaporation efficiency through a linear fit of its decrease \cite{Ketterle96_AAMOP}:
\begin{align}
    {\cal E}_{\rm evap}
    &=
    -\frac{ \partial \log_{10}\rho_{\rm PSD} }{ \partial \log_{10} N }. 
\end{align}
The PSD is obtained from temperature measurements of the simulation ensemble via the local density approximated relation 
\begin{align} \label{eq:PSD_vs_ToTF}
    {\rm Li}_3\left( \frac{ \rho_{\rm PSD} }{ \rho_{\rm PSD} - 1 } \right)
    &=
    -\frac{ 1 }{ 6 }
    \left(
    \frac{ T }{ T_F }
    \right)^{-3},
\end{align}
where ${\rm Li}_3(z)$ is the trilogarithmic function. 

To effectively simulate the evaporation trajectory following a lowering of the trap depth, we opt to use a position space cutoff scheme. 
That is, a molecule is taken as evaporated from the trap if it falls past the outer turning point along $z$, or goes past a position that is 6 times the thermal width of the initial cloud from the trap minimum:
\begin{subequations}
\begin{align} 
    \abs{ r_{\nu} - r_{\nu, \min} }
    &>
    6 \sqrt{ \frac{ k_B T_0 }{ m \omega_{\nu}^2(0) } }, \\
    \text{or}\quad  
    z < 1.2 z_{\max},
\end{align}
\end{subequations}
where $T_0$ is the initial equilibrium temperature, $r_{\nu, \min}$ is the position of the trap minimum along axis $\nu$ and $z_{\max} < 0$ is the trap local maxima along $z$. 
The criteria above allow us to account for the anisotropic molecular loss in space, as results from the gravitational trap sag seen in Fig.~\ref{fig:trap_potential_surfXZ}. 
Furthermore, the Gaussian trap profiles in the transverse directions imply that if molecules are too far away from the trap minima, they will no longer experience a large enough restorative potential, nor collisions, to return toward the trap center and are thus effectively evaporated away. 

We point out that while ${\cal E}_{\rm evap}$ provides a useful guide for experiments, it does not guarantee to achieve the highest phase space density amongst other schemes with possibly lower predicted efficiencies. For instance, a rapid decrease in the trap depth would still allow favorable evacuation of hot molecules and a seemingly efficient decrease in total energy.    
Unfortunately, the subsequent sample would have had no time to thermalize during the fast quench, disallowing the thermal tails of the distribution from being re-populated for further evaporative cooling beyond the initial evacuation.   
It is therefore useful to also track the final $T/T_F$ and $\rho_{\rm PSD}$ achieved, toward the goal of deeply degenerate Fermi gases.

\subsection{ In-simulation thermometry }

To mimic the experimentally extracted values of $T$, we utilize a Fermi-Dirac fit to the $y$-integrated simulation ensemble, likened to the optical density (OD) from absorption imaging of the molecular cloud \cite{DeMarco01_UCB, Regal05_PRL, Schindewolf22_Nat}:
\begin{align}
    {\rm OD}(x, z)
    &=
    \frac{ {\rm OD}_{\max} }{ {\rm Li}_2\left( -\zeta \right) }
    {\rm Li}_2\left( -\zeta e^{ -\frac{ x^2 }{ 2 \sigma_x^2 } - \frac{ z^2 }{ 2 \sigma_z^2 } } \right),
\end{align}
where ${\rm OD}_{\max}$ is the peak optical depth, $\sigma_i$ are the distribution widths, $\zeta$ is the fugacity and ${\rm Li}_2(z)$ is the dilogarithmic function.
In time-of-flight imaging, the distribution widths will evolve in time for every time-of-flight instance as
\begin{align}
    \sigma_{\nu} 
    =
    \frac{ \sqrt{ 1 + \omega_{\nu}^2 t_{\rm TOF}^2 } }{ \omega_{\nu} }
    \sqrt{ \frac{ k_B T_{\nu} }{ m } }, 
\end{align}
over the time interval $t_{\rm TOF}$. 
In the long time limit, this time-dependence changes the density images from position to momentum space distributions, since $r_{\nu} \rightarrow v_{\nu} t_{\rm TOF}$ and $\sigma_{\nu}  \rightarrow t_{\rm TOF} \sqrt{ k_B T / m }$, rendering
\begin{align}
    \left. {\rm OD}(x, z) \right|_{t_{\rm TOF} \rightarrow \infty}
    &=
    {\rm OD}(v_x, v_z) \\
    &=
    \frac{ {\rm OD}_{\max} }{ {\rm Li}_2\left( -\zeta \right) }
    {\rm Li}_2\left( -\zeta e^{ -\frac{ m ( v_x^2 + v_z^2 ) }{ 2 k_B T } } \right), \nonumber
\end{align}
where $v_{\nu}$ is the velocity along axis $\nu$. 

Leaving $T$ and $\zeta$ as float parameters, ${\rm OD}(v_x, v_z)$ is then fitted to the simulation distribution, obtained by constructing an appropriately normalized 2D histogram from the simulated particle ensemble, projected into the $x,z$ plane.   
In practice, obtaining the fugacity by fitting to the shape of the distribution results in large errors with noisy data. So we opt to utilize the relation 
in Eq.~(\ref{eq:PSD_vs_ToTF}) and $\zeta = \rho_{\rm PSD} (1 - \rho_{\rm PSD})^{-1}$
to infer the fugacity, floating only $T$.
If $T > T_F$, we simply revert to assuming a Boltzmann distributed gas, with temperature related to the mean-squared momenta $T = \langle \boldsymbol{p}^2 \rangle / (3 m k_B)$.

\section{ Numerical Results \label{sec:numerical_results} }

The simulation and measurement methods thus far described, are what we utilize to produce the data plotted in Fig.~\ref{fig:ToTF_vs_N_comparison}, providing us a positive benchmark against actual experimental data.  
Along with the parameters provided in Ref.~\cite{Schindewolf22_Nat} and $t_d = 2.5 \tau$, the close agreement was achieved by utilizing a 2-body loss rate constant of $\beta_L = 10^{-12}$ cm$^3$/s, and an added background heating rate of $\kappa = 100$ nK/s as reported by the experiment \footnote{
We suspected that background heating was caused by a sloshing mode excited in the gas upon sample preparation. 
}. 
The added background heating was simulated with momentum kicks during each simulation time step, taking the momentum of particle $k$, and increasing by $\boldsymbol{p}_k \rightarrow \boldsymbol{p}_k \left( 1 + 2 p_k^{-2} m k_B \kappa \Delta t \right)$, after the second Verlet integration step of Eq.~(\ref{eq:Verlet_integration}).

\begin{table}[ht]
\caption{ \label{tab:trap_parameter}
Table of parameter values for the potential confining a gas of fermionic $^{23}$Na$^{40}$K molecules. $h$ denotes Planck's constant. }
\begin{ruledtabular}
\begin{tabular}{l c c c}
    \multicolumn{1}{c}{\textrm{Parameter}} & \multicolumn{1}{c}{\textrm{Symbol}} & \multicolumn{1}{c}{\textrm{Value}} & \multicolumn{1}{c}{\textrm{Unit}} \\
    \colrule
    Beam 1 vertical width & $W_{1,z}$ & 57.5 & $\mu$m \\
    Beam 1 horizontal width & $W_{1,\perp}$ & 113 & $\mu$m \\
    Beam 1 wavelength & $\lambda_{1}$ & 1064 & nm \\
    Beam 1 power & $P_{1}$ & 0.242 & W \\
    Polarizability in beam 1 & $\alpha_{1}$ & $2.79 \times 10^{-3} h$ & m$^2$Hz/W \\
    Beam 2 vertical width & $W_{2,z}$ & 45 & $\mu$m \\
    Beam 2 horizontal width & $W_{2,\perp}$ & 156 & $\mu$m \\
    Beam 2 wavelength & $\lambda_{2}$ & 1064 & nm \\
    Beam 2 power & $P_{2}$ & 0.253 & W \\
    Polarizability in beam 2 & $\alpha_{2}$ & $2.79 \times 10^{-3} h$ & m$^2$J/W 
\end{tabular}
\end{ruledtabular}
\end{table}

\begin{figure}[ht]
    \centering
    \includegraphics[width=\columnwidth]{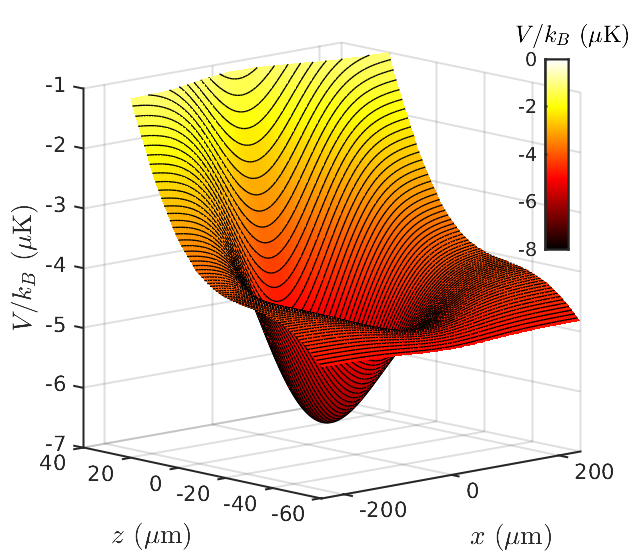}
    \caption{ The xODT potential energy surface $V(\boldsymbol{r})$, plotted as a function of coordinates $x$ and $z$ along $y = 0$, utilizing the trap parameters of Tab.~\ref{tab:trap_parameter}.  }
    \label{fig:trap_potential_surfXZ}
\end{figure}

Assurance of physically accurate simulations now motivates us to investigate tunable parameters for identifying efficient evaporation schemes. 
In this study, evaporation simulations commence with the default laser parameters listed in Tab.~\ref{tab:trap_parameter}, which results in the initial trap potential energy surface provided in Fig.~\ref{fig:trap_potential_surfXZ}. The plot is a slice along the $y = 0$ plane, showing a trap depth of $\approx 2.5$ $\mu$K. 
Lowering of the trapping beam power proceeds with $t_d = 2\tau$.

\subsection{ Dependence on 2-body loss \label{sec:2bodyloss} }

\begin{figure}[ht]
    \centering
    \includegraphics[width=\columnwidth]{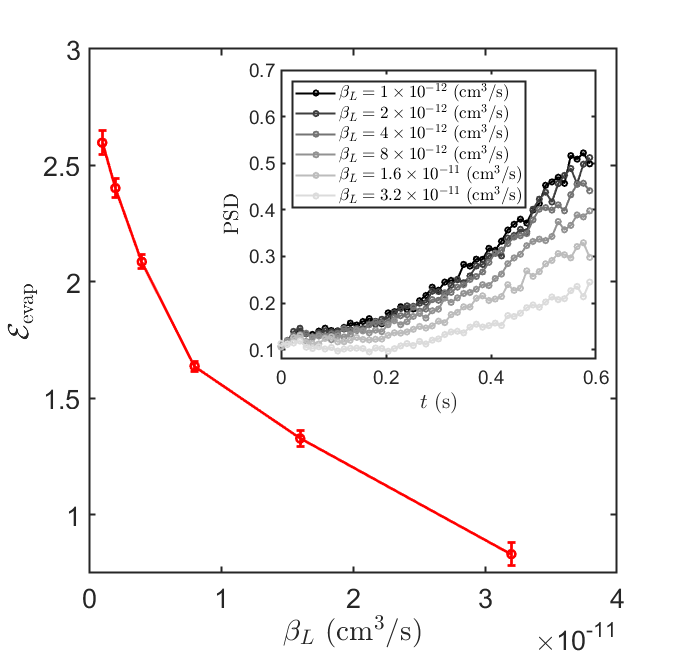}
    \caption{ Simulated evaporation efficiency ${\cal E}_{\rm evap}$ as a function of a constant 2-body loss rate $\beta_L$ (red data in main plot). Error bars are the linear fit uncertainties.   
    The inset gives the PSD trajectories as a function of time (gray scale curves) for each simulated value of $\beta_L$. All simulation data points (circles) are interpolated with solid lines to guide the eye. }
    \label{fig:evapPSDslope_vs_betaLossRate}
\end{figure}

A major factor that limits the efficiency of evaporating molecular samples to quantum degeneracy is 2-body collisional losses. Even with microwave shielding applied, inelastic collisions can still occur as a result of couplings between the dressed adiabatic channels. Occurring primarily in the region of higher density, such losses inevitably cause a flattening of the momentum distribution peak which results in antievaporative heating \cite{Ketterle96_AAMOP}. For a systematic study, we first explore the dependence on ${\cal E}_{\rm evap}$ as a function of temperature-independent 2-body loss rate values: $\beta_L = 1, 2, 4, 8, 16, 32$ ($10^{-12}$cm$^3$/s). 
Forced evaporation commences at $T = 1.1 T_F$ and occurs over $500$ ms, followed by a hold time of $100$ ms to allow the sample to thermalize. A low of $T \approx 0.6 T_F$ is reached in these simulations, placing the gas only weakly in the quantum degenerate regime \cite{Aikawa14_PRL}.   
As expected, we observe a trend of decreasing efficiency with increasing 2-body loss from our simulations as shown in Fig.~\ref{fig:evapPSDslope_vs_betaLossRate}.

In actuality, the 2-body loss rate has a temperature dependence inherited from the energy dependence of the integral inelastic cross section $\sigma_{\rm inel}$. 
For the typical temperatures of $\sim 200$ nK at which evaporation is expected to commence, the inelastic collision rate generally decreases with decreasing temperature. The trend of Fig.~\ref{fig:evapPSDslope_vs_betaLossRate} therefore serves as a theoretical worst case one might expect with 2-body loss, which we will use to explore evaporation in deeply degenerate samples later in the paper. 

Now extending our simulation to more faithful depictions of physical realizations, we incorporate inelastic scattering with $\sigma_{\rm inel}$ obtained from full scattering calculations.
For completeness, we briefly outline these calculations here, which closely follow the approaches detailed in Refs.~\cite{Karman18_PRL, Karman20_PRA, Karman22_PRA, Deng23_PRL}. We treat the NaK molecules as rigid rotors, considering only the lowest two rotational levels (a total of four rotational states) in our calculation. 
In the presence of a circularly polarized microwave field, all molecules are prepared in their highest dressed state. Consequently, we only need to consider the scattering of symmetrized two-molecule states, where it turns out that only seven of these are mutually coupled. To account for the short-range loss, we include an attractive van der Waals term $-C_6/r^6$, with $C_6 = 5 \times 10^5 ~\mathrm{a.u.}$~\cite{Yan20_PRL}, and a capture boundary condition~\cite{Light76_JCP, Clary87_FDCS, Rackham01_CPL} imposed at $r=50\,a_0$. 
Using the log-derivative method~\cite{Johnson73_JCP}, the scattering wavefunctions are numerically propagated to a large distance $r_M$ ($> 10^5\,a_0$) and then matched with asymptotic solutions to obtain the scattering $K$-matrix. The elastic and inelastic cross sections are then computed from the $K$-matrix.

We utilize up to $\ell = 11$ partial waves to ensure convergence of our scattering calculations. 
These calculations were performed at several well-chosen logarithmically spaced collision energies, from which we interpolate these values to construct a smooth function of $\sigma_{\rm inel}$ vs $E$ for Monte Carlo sampling. The incident angular dependence of inelastic scattering is not treated in this work.

\begin{figure}[ht]
    \centering
    \includegraphics[width=\columnwidth]{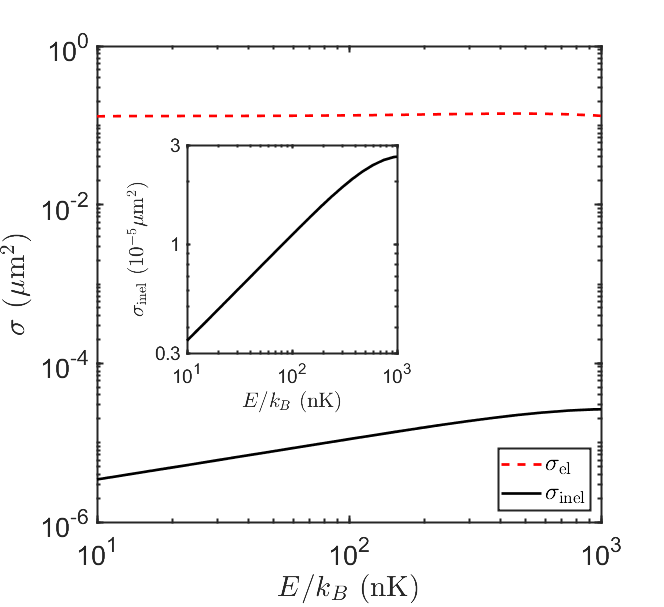}
    \caption{ Plot of the elastic (dashed red) and inelastic (solid black) cross sections as a function of collision energy on a log-log scale. The microwave Rabi frequency and detuning are both set at $50$ MHz.  }
    \label{fig:sigma_vs_E__OmegaDelta50MHz}
\end{figure}

Within the range of microwave parameters studied, the 2-body loss rate is shown to be exceptionally low with circularly polarized microwaves at $\Omega = 2\pi \times 50$ MHz and comparably large detuning. By setting $\Delta = 2\pi \times 50$ MHz, we find that $\sigma_{\rm inel}$ remains up to $4$ orders of magnitude smaller than $\sigma_{\rm el}$ as seen given in Fig.~\ref{fig:sigma_vs_E__OmegaDelta50MHz} (refer to Tab.~\ref{tab:dipole_parameters} for the microwave relevant parameters). 
For collision energies $\lesssim 200$ nK, the inelastic cross section is seen to scale as $\sqrt{ E }$, resulting in a 2-body loss rate that scales as $T$, consistent with $p$-wave dominated loss in identical fermions \cite{Ni10_Nat}.
Furthermore, $\sigma_{\rm el}$ only has a weak dependence on $E$, and so is well approximated by its energy-independent value at threshold. 
Favorably, we can infer from Fig.~\ref{fig:sigma_vs_E__OmegaDelta50MHz} that if evaporation commences at $T = 200$ nK, the initial 2-body loss rate constant is around $\beta_L \approx 2 \times 10^{-13}$, expected to allow for efficient evaporation with ${\cal E}_{\rm evap} > 2.5$ (see Fig.~\ref{fig:evapPSDslope_vs_betaLossRate}).  

\begin{table}[ht]
\caption{ \label{tab:dipole_parameters} 
Table of microwave parameters and resultant dipole scales. $a_0$ is the Bohr radius, $D$ is a Debye and $k_B$ is Boltzmann's constant. }
\begin{ruledtabular}
\begin{tabular}{l c c c}
    \multicolumn{1}{c}{\textrm{Parameter}} & \multicolumn{1}{c}{\textrm{Symbol}} & \multicolumn{1}{c}{\textrm{Value}} & \multicolumn{1}{c}{\textrm{Unit}} \\
    \colrule
    Microwave frequency & $\Omega$ & $2\pi \times 50$ & MHz \\
    Microwave detuning & $\Delta$ & $2\pi \times 50$ & MHz \\
    Effective dipole moment & $d$ & 0.56 & D \\
    Effective dipole length & $a_d$ & 2740 & $a_0$ \\
    Effective dipole energy & $E_{\rm dd}/k_B$ & 732 & nK 
\end{tabular}
\end{ruledtabular}
\end{table}

\subsection{ Trap sequences for efficient evaporation \label{sec:trap_sequences} }

Having fixed the microwave parameters, we look to study experimentally tunable trap parameters and their effect on ${\cal E}_{\rm evap}$. 
By varying the laser power, experiments can achieve various final potential depths at different rates.   
In practice, we take that the laser power of both xODT beams is lowered simultaneously, so that the ratio of final to initial trap power 
\begin{align}
    r_P 
    =
    \frac{ P_i(\tau) }{ P_i(0) },
\end{align}
changes equally for both beams $1$ and $2$, over a time interval $\tau$ with the time dependence in Eq.~(\ref{eq:power_time_dependence}).   
To obtain the functional dependence of ${\cal E}_{\rm evap}$ on these parameters, we run simulations for $\tau = 0.1, 0.5, 1, 1.5, 2, 2.5, 3$ s, and $p = 0.6$ to $0.9$ in steps of $0.05$.
In addition to forced evaporation, we also include a $100$ ms hold time after $\tau$ at the same trap depth. Although still resulting in some amount of plain evaporation, the added hold time allows the evaporated sample to further thermalize for more accurate thermometry.
A single-body molecular lifetime of 9 s is also included.

\begin{figure}[ht]
    \centering
    \includegraphics[width=\columnwidth]{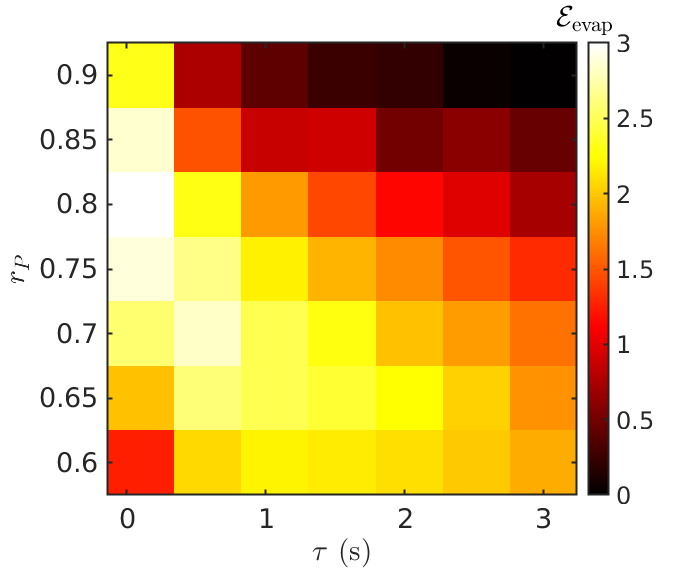}
    \vspace{-15pt}
    \caption{ The evaporation efficiency ${\cal E}_{\rm evap}$, as a function of the final relative power $r_P$ and evaporation time $\tau$. }
    \label{fig:evapPSDslope_vs_frp_vs_tevap}
\end{figure}

\begin{figure}[ht]
    \centering
    \includegraphics[width=\columnwidth]{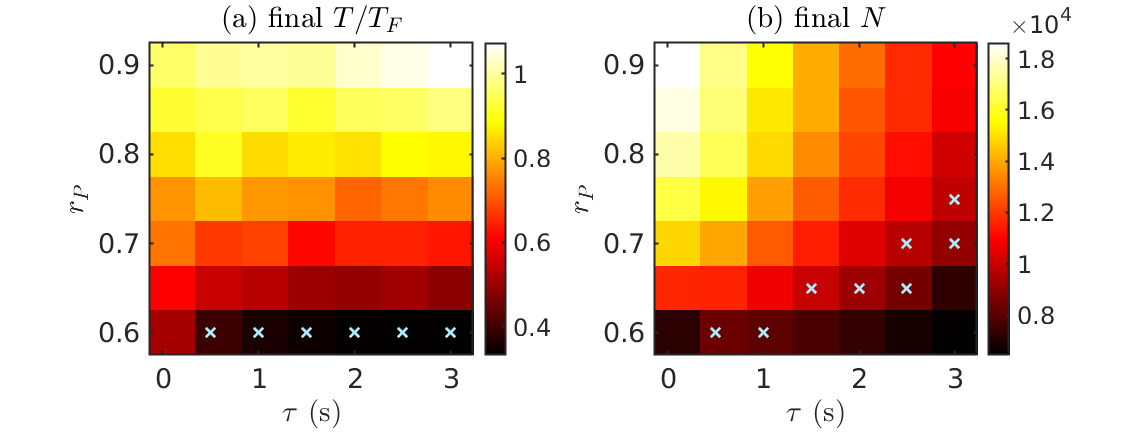}]
    \caption{ The (a) measured temperature ratio $T/T_F$ and (b) molecule number $N$ at the end of each evaporation trajectory, as a function of the in final relative power $r_P$ and evaporation time $\tau$. }
    \label{fig:finalNToTF_vs_frp_vs_tevap}
\end{figure}

Starting all numerical experiments with an initial temperature of $T = 200$ nK and $N = 20,000$ molecules ($T_F(0) \approx 175$ nK), the resulting variation of ${\cal E}_{\rm evap}(r_P, \tau)$ is visualized in Fig.~\ref{fig:evapPSDslope_vs_frp_vs_tevap}.
From this plot alone, a cursory glance indicates that rapid evaporation with $\tau = 0.1$ to a final relative power of $r_P = 0.8$ is optimal with regards to ${\cal E}_{\rm evap}$.  
But as alluded to at the start of this section (\ref{sec:evap_simulations}), this apparent gain in efficiency is not very useful in practice, only achieving a meager increment of $\rho_{\rm PSD}$ from $\sim 0.1$ ($T/T_F \approx 1.1$) to $0.2$ ($T/T_F \approx 0.85$) [see subplot (a) of Fig.~\ref{fig:finalNToTF_vs_frp_vs_tevap}, plotting $T/T_F$ as a function of $r_P$ and $\tau$].

In the hopes of achieving deeply degenerate gases, perhaps more appropriate is to first choose target parameters for the molecular sample. For instance, one might aim to achieve molecule numbers of $N > 8000$ and a final temperature of $T/T_F < 0.4$. These targets are indicated with light blue crosses in Fig.~\ref{fig:finalNToTF_vs_frp_vs_tevap}. Between the 2 common squares in subplots (a) and (b), Fig.~\ref{fig:evapPSDslope_vs_frp_vs_tevap} tells us that this final molecular sample is most efficiently achieved by setting $\tau = 1$ s and $r_P = 0.6$, with a predicted efficiency of ${\cal E}_{\rm evap} \approx 2.2$.  
For comparison, Bose-Einstein condensation of dipolar molecules was recently achieved with ${\cal E}_{\rm evap} \approx 2.0$ \cite{Bigagli24_Nat}, while atomic samples can reach much higher efficiencies of ${\cal E}_{\rm evap} \approx 3.5$ with their low 2-body losses \cite{Aikawa12_PRL}.

For a more experiment-agnostic guide, we can express the identified optimal evaporation time $\tau_{\rm opt}$ in terms of the inverse standard collision rate at the start of evaporation.
For elastic scattering, this is found to be $\tau_{\rm coll}(0) \approx 1.8$ ms, which gives the comparison $\tau_{\rm opt} \approx 555.6 \tau_{\rm coll}(0)$. 
As for the initial inelastic collision time, we find $\tau_{\rm inel}(0) = \langle n \sigma_{\rm inel} v_r \rangle^{-1} \approx 11.6$ s, granting us $\tau_{\rm opt} \approx 0.1 \tau_{\rm inel}(0)$.
So generally speaking, if evaporation can occur more than 10 times faster than inelastic collisions do, but around 550 times slower than elastic ones, evaporative cooling from $T / T_F \approx 1$ is slated to achieve deep Fermi degeneracy.

\onecolumngrid

\twocolumngrid

\section{ Prospects for evaporation to p-wave superfluidity \label{sec:superfluid_prospects} } 

Although a triumph for molecular experiments, temperatures of $T < 0.4 T_F$ have already been experimentally achieved in Ref.~\cite{Schindewolf22_Nat}, albeit with lower molecular numbers than those predicted as achievable here. Hence, a natural next step is to push the molecular gas into deeper quantum degenerate regimes. 
With $p$-wave superfluidity stipulated to occur at $T \lesssim 0.1 T_F$ \cite{Deng23_PRL}, we attempt a preliminary analysis with our evaporation simulator for achieving such temperatures with a microwave-shielded NaK gas.  
Even with perfect microwaves and arbitrarily stable lasers, the curse of inelastic collisions continues to plague the quest for colder molecular samples. For small $T/T_F$ in particular, Pauli blocking suppresses elastic thermalizing collisions \cite{Wang21_PRA} but not inelastic ones, since ultracold inelastic collisions lead to exothermic loss of molecules from the trap, outside which is void of a Fermi sea of molecules. The result is a marked decrease in the evaporation efficiency. 
To this end, we explore the possibilities for further evaporative cooling in low-temperature samples after the evaporation sequence in Sec.~\ref{sec:trap_sequences}.
In particular, we aim to find bounds on allowable 2-body loss rates where cooling through forced evaporation still overcomes antievaporative heating from inelastic loss.

\begin{figure}[b]
    \centering
    \includegraphics[width=\columnwidth]{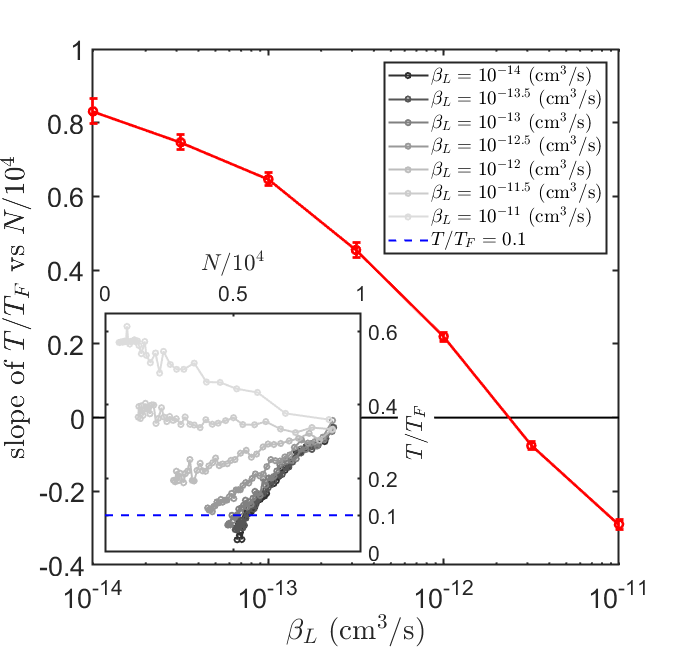}
    \caption{ The slope of $T/T_F$ vs $N/10^4$ for various values of $\beta_L$ (red data in main plot). Error bars are the linear fit uncertainties. The lower left inset shows the simulated $T/T_F$ vs $N/10^4$ trajectories during forced evaporation for $800$ ms, labeled by the upper right legend. All simulation data points (circles) are interpolated with solid lines to guide the eye. }
    \label{fig:slope_vs_2bodyLoss}
\end{figure}

Following an initial forced evaporation ramp that leaves the gas at $T = 0.35 T_F$ with $N = 9000$ molecules (refer to Fig.~\ref{fig:finalNToTF_vs_frp_vs_tevap}), we apply a secondary evaporation ramp from a trap depth of 150 nK to 80 nK in 800 ms. To systematically tune inelastic collisions, we once more adopt a temperature-independent two-body loss rate (see Sec.~\ref{sec:2bodyloss}) and vary it from $\beta_L = 10^{-14}$ to $\beta_L = 10^{-11}$ cm$^3$/s. The resulting slopes of $T/T_F$ vs $N$ for logarithmically-spaced $\beta_L$ values are plotted in Fig.~\ref{fig:slope_vs_2bodyLoss}, from which we find that $\beta_L \approx 2 \times 10^{-12}$ sets a break-even point below which further cooling can be achieved in spite of antievaporative heating. The actual $T/T_F$ vs $N$ time traces are given in the bottom left inset of Fig.~\ref{fig:slope_vs_2bodyLoss} (with a corresponding legend in the top right), showing that $T < 0.1 T_F$ is in fact achievable when $\beta_L \sim 10^{-13}$ cm$^3$/s.   
With the current trap and molecular parameters, the secondary evaporation ramp would commence at $T \approx 30$ nK, corresponding to a 2-body loss rate constant of around $\beta_L \approx 3 \times 10^{-14}$ cm$^3$/s (see Fig.~\ref{fig:sigma_vs_E__OmegaDelta50MHz}).  
As such, we conjecture that the $p$-wave superfluid phase is realizable in a 3D gas of microwave-shielded molecules by evaporative cooling.

A more decisive study must include 
the Hartree-Fock dipolar interaction terms that are expected to be significant at these deeply degenerate temperatures, which our current simulations do not. 
These effects are known to deform the Fermi gas in phase space \cite{Miyakawa08_PRA, Zhang09_PRA}, possibly leading to dynamical effects that could alter the evaporative cooling trajectories \cite{Sogo09_NJP, Baillie10_PRA}.
A pseudospectral scheme might be utilized to efficiently incorporate these effects \cite{Ronen06_PRA}, but we cater such inclusions to a future publication.  
Furthermore, at the low temperatures attained in the second ramp sequence, we found it difficult to satisfy the conditions of Eq.~(\ref{eq:smoothing_conditions}), so a bootstrap sampling of the in-simulation distribution to increase $N_{\rm sim}$ is done to maintain them (see App.~\ref{app:boostrap_sampling} for details).
Nevertheless, this work presents a pathway to understanding and achieving a strongly interacting molecular Fermi gas at unprecedented depths of quantum degeneracy.

\section{ \label{sec:conclusion} Outlook and Conclusions }

We have numerically studied evaporative cooling in a 3-dimensional gas of $^{23}$Na$^{40}$K molecules, made strongly dipolar and collisionally stable by applying circularly polarized microwaves. 
We employ a direct simulation Monte Carlo solver, which permits efficient sampling of both elastic and inelastic collisions amongst molecules, incorporating Pauli blocking effects due to fermionic quantum many-body statistics. 
Along with an accurate model of the trap potential, evaporation process and experimentally utilized thermometry, our simulation has shown favorable agreement with experimental data.

We then utilized our simulator to study optimal schemes for evaporation, primarily through varying the duration of forced evaporation and the final trap depth.
In doing so, we find that the evaporation efficiency on its own may not be a comprehensive metric for informing experiments attempting deeply degenerate samples.
Instead, we propose that the target thermodynamic state of the gas should also be considered as a constraint, over which the evaporation efficiency can be optimized to achieve it. 
Finally, we explored the possibilities of evaporative cooling in the deeply degenerate regime down to temperatures of $T \leq 0.1 T_F$. Our preliminary analysis shows promise for a molecular Fermi gas to reach this regime if two-body losses are sufficiently suppressed, although nonequilibrium Hartree-Fock dipolar effects have yet to be incorporated.   

We note that throughout this study, the initial molecule numbers and dipole moments tend to have evaporation occur close to or weakly in the hydrodynamic regime. Although a deeply hydrodynamic sample is expected to lower evaporation efficiency \cite{Ma03_JPB} due to hydrodynamic excitations \cite{Wang23_PRA, Wang23_PRA2}, we have found that evaporation can still reach efficiencies of up to ${\cal E}_{\rm evap} \gtrsim 2$ in our current regime, proven sufficient to achieve Bose-Einstein condensation of bosonic NaCs molecules \cite{Bigagli24_Nat}. 
Future works could explore the effect of performing evaporation from the dilute to hydrodynamic regimes and the dependence on microwave-induced dipolar interactions.

\begin{acknowledgments}

The authors acknowledge stimulating discussions with the \textit{NaK molecules} team at the Max-Planck-Institut f\"ur Quantenoptik. R.R.W.W thanks O. Goulko for helpful discussions on Pauli blocking.  
This material is based upon work supported by the National Science Foundation under Grant Number PHY 2317149. S. B., S. E., and X.-Y. L. acknowledge support from the Max Planck Society, the European Union (PASQuanS Grant No. 817482) and the Deutsche Forschungsgemeinschaft under German Excellence Strategy -- EXC-2111 -- 390814868 and under Grant No.\ FOR 2247.

\end{acknowledgments}

\appendix

\section{ The harmonic trap approximation \label{app:harmonic_trap} }

Around the trap minima, $V_{\rm ODT}$ can be expanded up to second order in spatial coordinates as 
\begin{align}
    V_{\rm harm}(\boldsymbol{r})
    &=
    \sum_{\nu} 
    A_{\nu} r_{\nu}^2,
\end{align}
dropping all constant and linear terms, which leaves the harmonic coefficients as 
\begin{subequations}
\begin{align}
    A_x 
    &=
    \frac{ \alpha_1 P_1 \lambda_1^2 \left( W_{1,y}^4 + W_{1,z}^4 \right) }{ \pi^3 W_{1,y}^5 W_{1,z}^5}
    +
    \frac{ 4 \alpha_2 P_2 }{ \pi W_{2,x}^3 W_{2,z} },
    \\
    A_y 
    &=
    \frac{ 4 \alpha_1 P_1 }{ \pi W_{1,y}^3 W_{1,z} }
    +
    \frac{ \alpha_2 P_2 \lambda_2^2 \left( W_{2,x}^4 + W_{2,z}^4 \right) }{ \pi^3 W_{2,x}^5 W_{2,z}^5},
    \\
    A_y 
    &=
    \frac{ 4 \alpha_1 P_1 }{ \pi W_{1,y}^3 W_{1,z} }
    +
    \frac{ 4 \alpha_2 P_2 }{ \pi W_{2,x}^3 W_{2,z} }.
\end{align}
\end{subequations}
This permits us to define harmonic trap frequencies as
\begin{align}
    \omega_{\nu}^2 
    =
    \frac{ 2 A_{\nu} }{ m }.
\end{align}

\section{ Trapping force \label{app:trapping_force} }

Given the trapping potential in Eq.~(\ref{eq:trap_potential}), the effective force felt by each molecule is given as $\boldsymbol{F} = -\grad V(\boldsymbol{r})$, which we compute in this section explicitly. For convenience, we further decompose $V_{\rm ODT}(\boldsymbol{r})$ by defining each cross propagating beam individually:
\begin{subequations}
\begin{align}
    V_{\rm ODT, 1}(\boldsymbol{r})
    &=
    - \frac{ 2 \alpha_1 P_1 
    \exp\left( 
    -\frac{ 2 y^2 }{ w_{1,y}^2(x) } 
    -
    \frac{ 2 z^2 }{ w_{1,z}^2(x) } 
    \right) }{ \pi w_{1,y}(x) w_{1,z}(x) }, \\
    V_{\rm ODT, 2}(\boldsymbol{r})
    &= 
    -\frac{ 2 \alpha_2 P_2 
    \exp\left( 
    -\frac{ 2 x^2 }{ w_{2,x}^2(y) } 
    -
    \frac{ 2 z^2 }{ w_{2,z}^2(y) } 
    \right) }{ \pi w_{2,x}(y) w_{2,z}(y) }.
\end{align}
\end{subequations}
Then taking the gradients of each term, we obtain
\begin{widetext} 
\begin{subequations} \label{eq:trap_force_gradients}
\begin{align} 
    -\frac{ \grad V_{\rm ODT,1}(\boldsymbol{r}) }{ V_{\rm ODT,1}(\boldsymbol{r}) }
    &=
    \begin{pmatrix}
        \frac{ 2 }{ x }
        + 
        \frac{ 4 \pi^4 }{ x }
        \left(
        \frac{ W_{1,y}^6 y^2 }{ ( \pi^2 W_{1,y}^4 + \lambda_1^2 x^2 )^2 }
        +
        \frac{ W_{1,z}^6 z^2 }{ ( \pi^2 W_{1,z}^4 + \lambda_1^2 x^2 )^2 }
        \right)
        -
        \frac{ \pi^2 }{ x }
        \left(
        \frac{ W_{1,y}^4 + 4 W_{1,y}^2 y^2 }{ \pi^2 W_{1,y}^4 + \lambda_1^2 x^2 }
        +
        \frac{ W_{1,z}^4 + 4 W_{1,z}^2 z^2 }{ \pi^2 W_{1,z}^4 + \lambda_1^2 x^2 }
        \right)
        \\
        \frac{ 4 \pi^2 W_{1,y}^2 y }{ \pi^2 W_{1,y}^4 + \lambda_1^2 x^2 } \\
        \frac{ 4 \pi^2 W_{1,z}^2 z }{ \pi^2 W_{1,z}^4 + \lambda_1^2 x^2 }
    \end{pmatrix}, \\
    -\frac{ \grad V_{\rm ODT,2}(\boldsymbol{r}) }{ V_{\rm ODT,2}(\boldsymbol{r}) }
    &=
    \begin{pmatrix}
        \frac{ 4 \pi^2 W_{2,x}^2 x }{ \pi^2 W_{2,x}^4 + \lambda_2^2 y^2 } \\
        \frac{ 2 }{ y }
        + 
        \frac{ 4 \pi^4 }{ y }
        \left(
        \frac{ W_{2,x}^6 x^2 }{ ( \pi^2 W_{2,x}^4 + \lambda_2^2 y^2 )^2 }
        +
        \frac{ W_{2,z}^6 z^2 }{ ( \pi^2 W_{2,z}^4 + \lambda_2^2 y^2 )^2 }
        \right)
        -
        \frac{ \pi^2 }{ y }
        \left(
        \frac{ W_{2,x}^4 + 4 W_{2,x}^2 x^2 }{ \pi^2 W_{2,x}^4 + \lambda_2^2 y^2 }
        +
        \frac{ W_{2,z}^4 + 4 W_{2,z}^2 z^2 }{ \pi^2 W_{2,z}^4 + \lambda_2^2 y^2 }
        \right)
        \\
        \frac{ 4 \pi^2 W_{2,z}^2 z }{ \pi^2 W_{2,z}^4 + \lambda_2^2 y^2 }
    \end{pmatrix}, \\ 
    -\grad V_{\rm g}(\boldsymbol{r})
    &=
    - m g.
\end{align}
\end{subequations}
Although the expressions above are algebraically non-divergent, a linear coordinate in the denominator for the first 3 gradients above might result in numerical instabilities. As such, we also present the first-order Taylor expansion with respect to the unstable coordinate in the relevant vector components:
\begin{subequations}
\begin{align}
    \left[ 
    \grad V_{\rm ODT,1}(\boldsymbol{r})
    \right]_x
    &\approx
    \frac{ 2 \alpha_1 P_1 }{ \pi W_{1,y} W_{1,z} }
    \left( 
    \frac{ 1 }{ W_{1,y}^4 }
    -
    \frac{ 4 y^2  }{ W_{1,y}^6 }
    +
    \frac{ ( W_{1,z}^2 - 4 z^2 ) }{ W_{1,z}^6 }
    \right)
    \frac{ \lambda_1^2 }{ \pi^2 }
    \exp(
    -\frac{ 2 y^2 }{ W_{1,y}^2 } 
    -
    \frac{ 2 z^2 }{ W_{1,z}^2 }
    ) x
    +
    {\cal O}( x^2 ), \\
    \left[
    \grad V_{\rm ODT,2}(\boldsymbol{r})
    \right]_y
    &\approx
    \frac{ 2 \alpha_2 P_2 }{ \pi W_{2,x} W_{2,z} }
    \left( 
    \frac{ 1 }{ W_{2,x}^4 }
    -
    \frac{ 4 x^2 }{ W_{2,x}^6 }
    +
    \frac{ ( W_{2,z}^2 - 4 z^2 ) }{ W_{2,z}^6 }
    \right)
    \frac{ \lambda_2^2 }{ \pi^2 }
    \exp(
    -\frac{ 2 x^2 }{ W_{2,x}^2 } 
    -
    \frac{ 2 z^2 }{ W_{2,z}^2 }
    ) y
    +
    {\cal O}( y^2 ). 
\end{align}
\end{subequations}
We utilize these linearized gradients instead of those in Eq.~(\ref{eq:trap_force_gradients}), for close-to-zero values of the respective coordinates in Monte Carlo simulations.
\end{widetext}

\section{ Bootstrap sampling from the in-simulation ensemble \label{app:boostrap_sampling} }

The conditions in Eq.~(\ref{eq:smoothing_conditions}) are difficult to satisfy when the simulation reaches temperatures of $T \lesssim 0.3 T_F$, but can be adequately accommodated by increasing $N_{\rm sim}$ [see (\ref{subeq:smoothing_lowerbound})].  
However, increasing $N_{\rm sim}$ from the start of evaporation can be computationally expensive and thus impractical. We work around this issue by increasing $N_{\rm sim}$ only when the (\ref{eq:smoothing_conditions}) conditions are no longer met, occurring after $N$ has already been greatly reduced from evaporation. 

We increase $N_{\rm sim}$ to satisfy (\ref{eq:smoothing_conditions}) by sampling from the energy distribution $f_E(E)$, inferred from the in-simulation ensemble, where $E(\boldsymbol{r}, \boldsymbol{p}) = \boldsymbol{p}^2/(2 m) + V(\boldsymbol{r})$ is the single-particle energy. 
By constructing a histogram with the simulated particle energies over an appropriately chosen energy grid, we interpolate this discrete distribution with a Gaussian process (GP) model \cite{Rasmussen05_MIT} to obtain an approximate but continuous representation of $f_E(E)$. We utilize a GP model trained with a Mat\'ern-$\frac{5}{2}$ kernel to prevent overfitting the histogram's statistical fluctuations from Monte Carlo sampling noise. 
New simulation particles are then sampled from the GP model of $f_E(E)$ until (\ref{eq:smoothing_conditions}) are satisfied once more, following which numerical time-evolution is resumed. 
Utilizing the energy distribution for bootstrap sampling of new simulation particles implicitly invokes the assumption of ergodicity \cite{Holland96_QSO}, which should be well satisfied with the large collisional cross sections between microwave-shielded NaK molecules.
Importantly, the parameter $\xi$ must be updated after the bootstrap sampling to preserve physically accurate collision rates. 

One might be concerned that having $N_{\rm sim}$ exceed $N$ causes simulation particle loss to no longer be faithful to actual molecular loss. However, we remind the reader that simulation particles represent discretized segments of the molecular distribution in phase space, and are merely a means for approximating the continuous profile of $f(\boldsymbol{r}, \boldsymbol{p})$. Evaporation with $N_{\rm sim} \neq N$ is, therefore, still a consistent approximation of Boltzmann equation dynamics.

\bibliography{main.bib} 

\end{document}